\documentclass[a4paper]{jpconf}
\usepackage{graphicx}
\usepackage{amsmath,amssymb,boldmath}
\usepackage[latin1]{inputenc}
\usepackage{colortbl}
\usepackage[english]{babel}
\usepackage{subfig}
\usepackage{wrapfig}
\usepackage{psboxit}
\usepackage{rotating}
\usepackage{times}
\usepackage{color}
\usepackage{pifont}
\usepackage{pstricks}
\usepackage{picinpar}
\usepackage{xspace}
\usepackage{latexsym}
\usepackage{moreverb}
\usepackage{here}
\begin{document}
\title{3-loop Feynman Integral Extrapolations for the Baseball Diagram
}

\author{E de Doncker$^1$,\ F Yuasa$^2$,\ T Ishikawa$^2$ and K Kato$^3$}

\address{$^1~$Department of Computer Science, Western Michigan University, Kalamazoo MI 49008, U.~S.~A.}

\address{$^2~$High Energy Accelerator Research Organization (KEK), Oho 1-1, Tsukuba, Ibaraki, 305-0801, Japan}

\address{$^3~$Department of Physics, Kogakuin University,
Shinjuku, Tokyo 163-8677, Japan}

\ead{elise.dedoncker@wmich.edu,\,fukuko.yuasa@kek.jp,\,tadashi.ishikawa@kek.jp,\\{\black katok@kute.tokyo}}

\begin{abstract}
We focus on numerical techniques for expanding 3-loop Feynman integrals
with respect to the dimensional regularization parameter $\varepsilon,$
which is related to the space-time dimension as $\nu = 4-2\varepsilon,$
and describes underlying UV singularities located at the boundaries of
the integration domain.
As a function of the squared momentum $s,$ the expansion coefficients
exhibit thresholds that generally delineate regions
for their computational techniques. 
For the problem at hand, a sequence 
of integrations with a linear extrapolation as $\varepsilon\rightarrow 0$
may be performed to determine leading coefficients of the $\varepsilon$-expansion numerically. 
For the ``baseball" Feynman diagram, we have used extrapolation with respect to an additional parameter
to improve the accuracy of the $\varepsilon$-expansion coefficients
in case of singularities internal to the domain.
\end{abstract}

\section{Introduction and Background}
{\color {black}
Higher order corrections are required for accurate theoretical 
{\color{black} prediction} 
improvements in the technology of high energy physics experiments.
Feynman loop integrals arise in the calculations of the Feynman 
diagrammatic approach, which is commonly used to address
higher order corrections.
Loop integrals may suffer from integrand singularities
or irregularities at the boundaries and/or in the interior
of the integration domain.
}

We recently explored methods for higher-order corrections to
2-loop Feynman integrals in the Euclidean or physical kinematical
region~\cite{talk-ACAT2022}, using numerical extrapolation and adaptive iterated
integration. Our current goal is to address a 3-loop two-point
integral for the ``baseball" diagram of Figure~1 with six internal lines.

A representation of an $L$-loop Feynman integral with $N$ internal lines is 
given by {\color {black} (e.g.,~\cite{nakanishi57,kinoshita74A,kinoshita74B,kinoshita74C})}
\begin{align}
& {\mathcal F} = {\Gamma(N-\frac{\nu L}{2})}\,
(-1)^N \int_{{\mathcal C}_N}\prod_{r=1}^{N}dx_{r}\, \delta(1-\sum x_{r}) \,U^{-\nu/2} (V-i\varrho)^{\nu L/2-N} \\ 
\label{Lloop}
& ~~~~~\mbox{or,~for~} L = 3, N = 6: \nonumber \\
& {\mathcal F} = {\Gamma(3\varepsilon)}\,
\int_{{\mathcal C}_6}\prod_{r=1}^{6}dx_{r}\, \delta(1-\sum x_{r}) \,U^{~\varepsilon-2} (V-i\varrho)^{-3\varepsilon} \\
\nonumber
\end{align}
where $V = M^2-W/U, ~~M^2 = \sum_r m_r^2 x_r$; ~
$U$ and $W$ are polynomials determined by the topology of the corresponding diagram and physical parameters;~ $\nu = 4-2\varepsilon$ is the space-time dimension; 
$\varrho$ is a regularization parameter and we can set $\varrho = 0$ unless $V$ vanishes in the domain;~
${\mathcal C}_N $ is the $N$-dimensional unit hypercube. 
We consider the integral without the prefactor {\color {black} $\Gamma(3 \varepsilon).$}

We use automatic integration, which is a black-box approach for
producing (as outputs) an approximation $Qf$ to an integral
$\mathcal I = \int_{\mathcal D} f(\vec{x}) ~d\vec{x}$
and an estimate ${\mathcal E}\hspace*{-0.5mm}f$ of the actual error $Ef = |Qf-\mathcal I|,$ with the goal of satisfying
an accuracy requirement of the form
$|\,Qf-\mathcal I\,| ~\le ~{\mathcal E}\hspace*{-0.5mm}f ~\le~ \max\,\{\,t_a\,,\,t_r\,| \mathcal I |\,\},$
where the integrand function $f,$ $d$-dimensional region $\mathcal D$ and (absolute/relative)
error tolerances $t_a$ and $t_r,$ respectively, are specified as part of the input.

For ultra-violet (UV) singularities at the boundaries of the domain, we make use of
lattice rules for Quasi-Monte Carlo (QMC) integration~\cite{sloan} 
with a boundary transformation that is capable of smoothing
singularities~\cite{sidi03,sidi06}. 
Another non-adaptive approach is based on a double-exponential 
(DE) formula~\cite{mori78,takahasi74,sugihara97,iccsa22}.
Adaptive integration may be useful in fairly low dimensions
to treat interior singularities by 
intensive region partitioning around hot spots~\cite{pi83,parintweb,csci19}. 
We give results obtained with 1D iterated adaptive integration by the
program {\sc Dqagse} from the {\sc Quadpack} package~\cite{pi83,jocs11}.
In~\cite{talk-ACAT2022} we applied adaptive integration with an extrapolation on the 
parameter $\varrho$ that adjusts the factor in $V$ in the integrand
denominator of~Eq.\,(1). We termed the combined extrapolations in $\varepsilon$
and $\varrho$ as a ``double extrapolation."

Subsequently, Section~2 defines the baseball integral; linear extrapolation and double extrapolation methods are
covered in Section~3; results are given in Section~4 and conclusions in
Section~5.

\section{Baseball integral}
\begin{wrapfigure}{R}{4cm}
\includegraphics[width=1.00\linewidth]{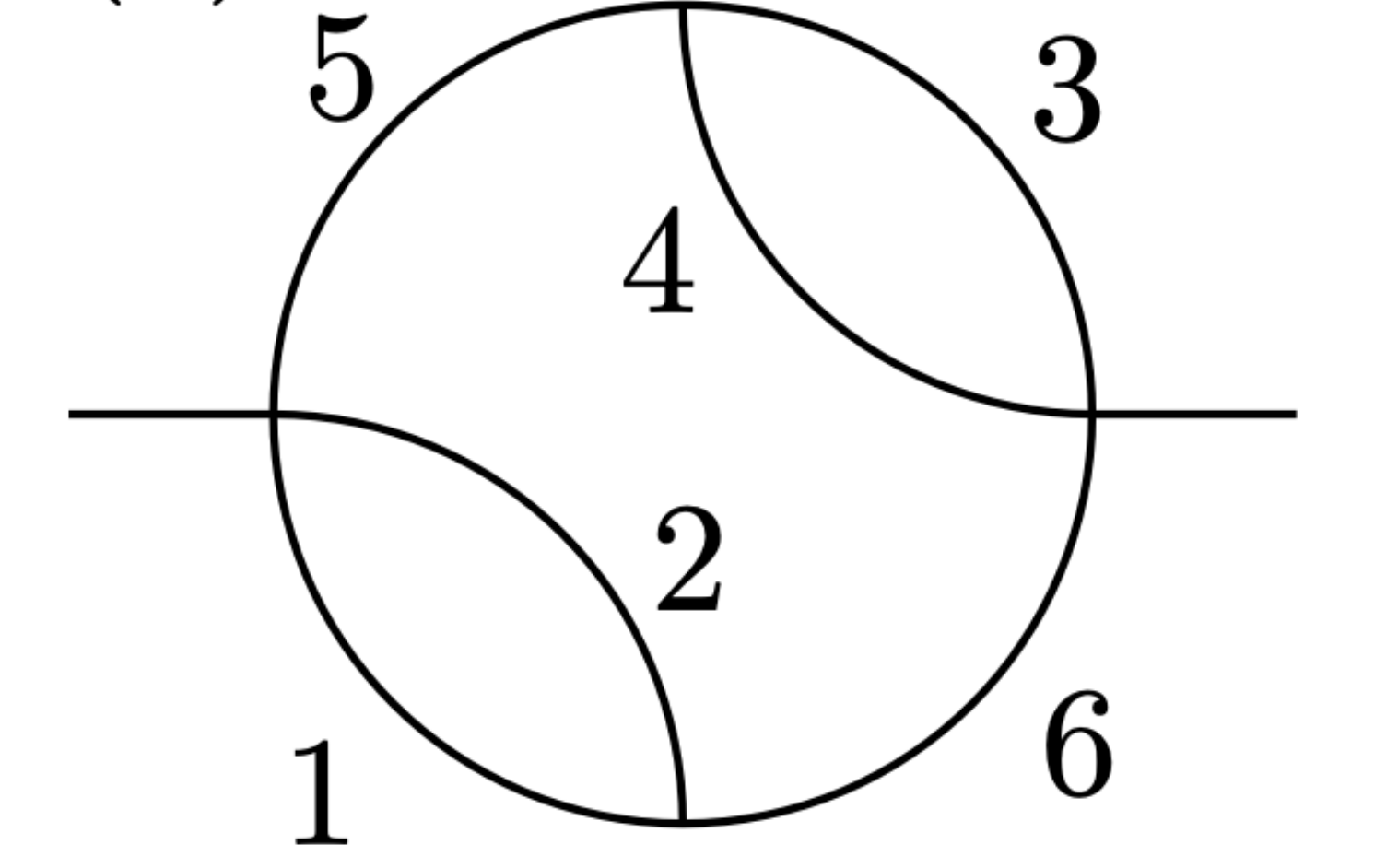}
\caption{3-loop \emph{baseball} diagram with $N = 6$ lines
}
\end{wrapfigure}
A set of 3-loop diagrams with six internal lines, and their
associated integrals are calculated by the authors
after separating the ultra-violet divergence{\color {black} ~\cite{ptep24a}}.
{\color {black} It is our goal to demonstrate our techniques for numerical integration and extrapolation using another  
3-loop diagram in the massive case. Hereafter, we
assume that all masses of the internal lines equal 1.
The derivation is based on a sector decomposition in
5D parameter space; thus the integrals are labeled by permutations
of the 5-tuple $(1,2,3,4,5).$ 
{\color {black} There are $5!=120$ integrals, whereas many of the integrals may coincide through symmetries and
the number of integrals to be computed is reduced.} 
We deal with the baseball diagram shown in Figure 1.
Then after assuming equal masses, 15 integrals remain.
{\color {black} In this paper, we will study the ${\mathcal I}_{51234}$ integral {\color{black}of} the baseball diagram, which {\color{black}is the integral that has the most singular UV divergence among the 15, with an} ${\mathcal O}(\frac{1}{\varepsilon^2})$ {\color {black} term,}
{\color {black} and} the rest start {\color{black}at the} ${\mathcal O}(\frac{1}{\varepsilon})$ or ${\mathcal O}(1)$ {\color{black}order}.} 

The integral is given by
\begin{eqnarray}
{\mathcal I_{51234}} &=& \int d\Gamma_X ~t^{-1+2\varepsilon}  v^{-1+\varepsilon} u^{\varepsilon}\, G^{-3 \varepsilon} f^{-2+\varepsilon} = I_{00} + I_{01} + I_{10} + I_{11} \\
I_{00} &=& \frac{1}{\varepsilon^2} \frac{1}{2} \int d\Gamma_0 ~u^{\varepsilon} G_{0}^{-3 \varepsilon} f_{0}^{-2+\varepsilon} \\
I_{01} &=& \frac{1}{\varepsilon} \frac{1}{2} \int d\Gamma_a ~v^{-1+\varepsilon} u^{\varepsilon}
\left(G_{a}^{-3 \varepsilon} f_{a}^{-2+\varepsilon} - G_{0}^{-3 \varepsilon} f_{0}^{-2+\varepsilon}\right)
\label{i01} \\
I_{10} &=& \frac{1}{\varepsilon} \int d\Gamma_b ~t^{-1+2\varepsilon} u^{\varepsilon}
\left(G_{b}^{-3 \varepsilon} f_{b}^{-2+\varepsilon} - G_{0}^{-3 \varepsilon} f_{0}^{-2+\varepsilon}\right) \\
I_{11} &=&  \int d\Gamma_X ~t^{-1+2\varepsilon} v^{-1+\varepsilon} u^{\varepsilon} \times \nonumber \\
& &\left(G_{}^{-3 \varepsilon} f_{}^{-2+\varepsilon} - G_{a}^{-3 \varepsilon} f_{a}^{-2+\varepsilon}
- G_{b}^{-3 \varepsilon} f_{b}^{-2+\varepsilon} + G_{0}^{-3 \varepsilon} f_{0}^{-2+\varepsilon}\right) \nonumber \\
&=& \int d\Gamma_X ~t^{-1+2\varepsilon} v^{-1+\varepsilon} u^{\varepsilon} \left(G_{}^{-3 \varepsilon} f_{}^{-2+\varepsilon} - G_{b}^{-3 \varepsilon} f_{b}^{-2+\varepsilon} \right) 
\label{i11sim}
\label{baseball}
\end{eqnarray}
where
\begin{eqnarray}
f &=& 1 + w + u + u w + t u + t u w + t u v w + t u^2 v w \label{ff} \\
f_0 &=& f|_{t=0,v=0}, ~~f_a=f|_{t=0}, ~~f_b=f|_{v=0} \\
q &=&z (1-z) + w z (1-z) + u z (1-z) + u w z (1-z) + t u z + t u w z + t u v w (1-z) \nonumber \\
  & &+ t u^2 v w (1-z) + t^2 u^2 v w \nonumber
\end{eqnarray}
\vspace*{-0.8cm}
\begin{eqnarray}
q_0 &=& q|_{t=0,v=0}, ~~q_a=q|_{t=0}, ~~q_b=q|_{v=0} \\
G   &=& (1+t(1+u(1+v(1+w))))-s\frac{q}{f} \\
G_0 &=& G|_{t=0,v=0}, ~~G_a=G|_{t=0}, ~~G_b=G|_{v=0} \label{G0} \\ \nonumber
\end{eqnarray}
\vspace*{-0.5cm}
and
\begin{align}
&\int d\Gamma_X = \int_0^1 dz  \int_0^1 dt  \int_0^1 du  \int_0^1 dv  \int_0^1 dw, ~~~~~~ \int d\Gamma_0 = \int_0^1 dz  \int_0^1 du  \int_0^1 dw \nonumber \\ 
&\int d\Gamma_a = \int_0^1 dz  \int_0^1 du  \int_0^1 dv  \int_0^1 dw, ~~~~~~~ \int d\Gamma_b = \int_0^1 dz  \int_0^1 dt  \int_0^1 du  \int_0^1 dw \nonumber \\ \nonumber
\end{align}
Note that $f_0 = f_a,$ $q_0 = q_a$ and $G_0 = G_a;$ thus the integrand of~Eq.\,(5) is zero,
and $I_{11}$ simplifies to~Eq.\,\eqref{i11sim}.
{\color {black} Here, $s$ equals $s=P^2$, where $P$ is the external momentum;
$s$ is measured in units of $m^2$.}

\section{Methods}
\subsection{Integration rules}
The double-exponential method (DE)~\cite{takahasi74,sugihara97,iccsa22}
transforms the one-dimensional integral
~$\int_{0}^{1} f(x) \,dx =\int_{-\infty}^{\infty} f\,(\phi(t))\,\phi'(t)\,dt$~
using
$x = \phi\,(t)=\frac{1}{2}(\rm{tanh}\,(\frac{\pi}{2}\rm{sinh}\,(t))+1),$ with $\phi'(t)=\frac{\pi\,\rm{cosh}\,(t)}{4\,\rm{cosh}^2(\frac{\pi}{2}\rm{sinh}\,(t))}.$ \\
DE involves a truncation of the infinite range
and an iterated application of the trapezoidal rule.

For QMC we apply a lattice rule with $\approx 10$M points,
for which we previously computed the generators (see, e.g.,~\cite{almulihi17})
using the component-by-component (CBC) algorithm~\cite{nuyens06,nuyens06a}.
For increased accuracy, we apply $m^d$-copy rules
$Q^{(m)}$ of a basic rank-1 lattice
rule~\cite{sloan} $Q$ with $m = 2$,
which corresponds to subdividing the
$[0,1]$-range into $m$ equal parts
in each coordinate direction and scaling the basic
rule to each subcube. An error estimate is also given.
The regular nature of the rule allows for an efficient implementation
on GPUs.

\subsection{Linear extrapolation and double extrapolation methods}
Linear extrapolation for an integral $\mathcal I$ is based on an asymptotic expansion of the form
\begin{align}
&{\mathcal I}(\varepsilon) \sim \sum_{k\ge \kappa} C_k \,\varphi_k(\varepsilon), ~~~~~~\mbox{as~~} \varepsilon \rightarrow 0 \nonumber
\end{align}
where the sequence of $\varphi_k(\varepsilon)$ is known. 
For the integrals at hand, $\kappa = -2$ and $\varphi_k(\varepsilon) = \varepsilon^k$ for $I_{00},$ $\kappa = -1$ for $I_{10},$ and $\kappa = 0$ for $I_{11}.$ 
The expansion is truncated after $2, 3, \ldots, n$ terms to form
linear systems
of increasing size in the $C_k$ variables. This is a generalized form of Richardson
extrapolation~\cite{brezinski80,sidi03}.

For fixed $\varepsilon = \varepsilon_\ell,$ the integrand 
may have a vanishing denominator in the interior of the
domain, say due to the factor $V^{-3\varepsilon}$ in Eq.\,(2).
Then $V$ can be replaced by $V-i\varrho.$
Since the structure of an expansion in the parameter $\varrho$ is unknown,
we apply a nonlinear extrapolation with the
$\epsilon$-algorithm~\cite{wynn56}
to a sequence of ${\mathcal I}(\varepsilon_\ell,\varrho)$ as
$\varrho \rightarrow 0$. 
The combined $\varepsilon$ and $\varrho$ extrapolations constitute
a double extrapolation{\color {black}~\cite{cpp10,acat11,yuasa-JPS22}}.

As a simple example, it can be seen that the integrand of $I_{00}$ in
Eq.\,(4) is singular within the domain ${\mathcal C}_3$ when $s \ge 4.$ 
Following Eqs.\,\eqref{ff}-\eqref{G0}, we have that
\begin{align}
& f_0 = 1+w+u+uw > 0 \nonumber \\
& q_0 = z(1-z)(1+w+u+uw) = z(1-z)f_0 \nonumber \\
& G_0 = 1-s\frac{q_0}{f_0} = 1-sz(1-z) = sz^2-sz+1 \nonumber
\end{align}
Thus, $G_0 = sz^2-sz+1 = 0$ has real solutions when the discriminant $D = s^2-4s = s(s-4) \ge 0,$
and $z = (s\pm \sqrt{s^2-4s})/(2s) = \frac{1}{2}\pm \frac{1}{2}\sqrt{\frac{s-4}{s}} = \frac{1}{2}(1\pm\sqrt{\frac{s-4}{s}}).$

Nevertheless, in view of the weak nature of the singularity,
it is possible to obtain solutions for $s \ge 4$ using single
extrapolations. However, in Section~\ref{results}, we illustrate 
a case where double extrapolation achieves more accuracy than single extrapolation {\color {black} with regard to $\varepsilon$}.

\section{Numerical results}
\label{results}
\begin{center}
\begin{table}[t]
\begin{scriptsize}
\caption{Results for $s = 1$ and $s = 4$ with QMC and $\varepsilon$-extrapolation}
\begin{center}
\begin{tabular}{|c|l|l|l|l|}\hline
$s = 1$ & $C_{-2}$ & $C_{-1}$ & $C_0$ & $C_1$\\ \hline
$I_{00}$& 0.125 & -0.026748351868&0.1229976127 &-0.087710974 \\ \hline 
NI & 0.124999999999962 & -0.026748351855 & 0.1229976114 & -0.087710933 \\ \hline \hline
$I_{10}$ &-  &-0.10136627702704 &-0.5057851162 &1.89367438 \\ \hline
NI & & -0.10136627702762 & -0.5057851170 & 1.89367429 \\ \hline \hline 
$I_{11}$&-  &- &-0.03011542756 &-0.045488315  \\ \hline 
NI & - & - & -0.03011542740 & -0.045488308 \\ \hline \hline
sum&0.125 &-0.128114628895 &-0.4129029311 &1.760475099 \\ \hline
sum NI&0.124999999999962 &-0.128114628883 & -0.4129029330 & 1.760475139 \\ \hline
\end{tabular}
\end{center}

\begin{center}
\begin{tabular}{|c|l|l|l|l|}\hline
$s = 4$ & $C_{-2}$ & $C_{-1}$ & $C_0$ & $C_1$\\ \hline
$I_{00}$&0.125 &0.65342641 &4.073701 &24.2815 \\ \hline 
NI & 0.1250000000010 & 0.65342699 & 4.073760 & 24.2839 \\ \hline \hline 
$I_{10}$&- &-0.1013662770270 &-4.003001 &-21.4910 \\ \hline 
NI & & -0.1013662770283 & -4.002911 & -21.4873 \\ \hline \hline 
$I_{11}$&-  &- &-0.030115427558 & -0.415010  \\ \hline 
NI & - & - & -0.030115427509 & -0.415361 \\ \hline \hline
sum&0.125 &0.55206013 &0.040585 &2.375546 \\ \hline
sum NI&0.1250000000010 & 0.55206071 & 0.040733 & 2.381239 \\ \hline
\end{tabular}
\end{center}
\end{scriptsize}
\end{table}

\end{center}
Results for $s=1$ and $s=4$ are summarized in Table~1, showing
leading coefficients of the Laurent series expansion of the integrals
$I_{00}, I_{!0}$ and $I_{11}.$ 
The integration uses a lattice rule with $~ 10M$ points, with two copies in each coordinate direction, and the Sidi $\Psi_6$ transformation~\cite{sidi06}.
The expansion with respect to $\varepsilon$ is performed with a linear (single) extrapolation.
``NI" refers to our ``Numerical Integration."

The ``exact" values listed for comparison were obtained
analytically and with
the Mathematica functions {\tt Integrate} and {\tt NIntegrate}.
Analytic exact values are given for $C_{-2},$ $C_{-1}$ and $C_0,$ 
and numerical values for $C_1.$
The $C_{-2}$ term appears only in $I_{00}$ and is $s$-independent. 
Similarly, $C_{-1}$ of $I_{10}$ is $s$-independent. 
When the Mathematica {\tt Integrate} function fails at the evaluation of 
$C_0$ and $C_1$ for $I_{10}$ and $I_{11}$ for $s = 1$ and 4, {\tt NIntegrate} is used. 
The lattice rule integrations were carried out on an Nvidia Quadro GV100 GPU,
each taking time of a fraction
of a second for about 20 integrations per problem, although less were used.

\begin{table}
\begin{center}
\begin{scriptsize}
\caption{Above threshold results with iterated adaptive integration and double extrapolation for $I_{00}$}
\begin{tabular}{|c|l|l|l|l|c|}\hline
s & $C_{-2}$ & $C_{-1}$ & $C_0$ & $C_1$ &\hspace*{-1.5mm} \#$\varepsilon$-ext.\\
  & & & & & T[s]\\ \hline
4.5 & 0.1249999983 & 0.56678339 & 0.4463330 & -6.00525 & 8 \\
    & $\pm 8.8$e-9 & $\pm 1.3$e-6 & $\pm 8.0$e-5 & $\pm 2.5$e-3 & 33 \\ \hline
5 & 0.1249999999989 & 0.4920230573484 & -0.74245949112 & -7.754615366 & 10  \\
  & $\pm 7.9$e-14 & $\pm 2.5$e-11 & $\pm 3.1$e-9 & $\pm 2.1$e-7 & 46 \\ \hline
7 & 0.124999999999963 & 0.2687848328350 & -2.744283992 & -4.54297095 & 12 \\
  & $\pm 7.4$e-14& $\pm 4.7$e-11 & $\pm 1.2$e-8 & $\pm 1.7$e-6 & 61 \\ \hline
10 & 0.12500000000016 & 0.054052104327 & -3.6553766977 & 1.92828020 & 12 \\
   & $\pm 3.5$e-13 & $\pm 2.2$e-10 & $\pm 5.8$e-8 & $\pm 8.1$e-6 & 66 \\ \hline
\end{tabular}
\end{scriptsize}
\end{center}
\end{table}

Sample results using double extrapolations are further given in Table~2 for $I_{00}$ with $s = 4.5, \,5, \,7$ and 10. 
The $\varrho$ values for the extrapolation follow a geometric sequence $\varrho_j = 2^{-j}$ for $j = 1, 2,\ldots, 11,$ 
whereas the $\varepsilon$ values form a Bulirsch type sequence~\cite{bulirsch64}, $\varepsilon_\ell = 1/4, 1/6, 1/8, 1/12, 1/16, 1/24, \ldots.$ 
The $\varepsilon$ results {\color {black} listed in Table~2} are incurred when the
estimated error (based on successive differences) no longer
decreases. 
$\varepsilon$-ext {\color {black} denotes} the number of $\varepsilon$ extrapolations
to obtain the result for this value of $s.$ 
T[s] denotes the corresponding time (for integration) incurred
through this number of $\varepsilon$-ext (not including the
iteration where the accuracy stops improving). The integration time
dominates the overall time.
The larger values of $s$ require more extrapolations for
convergence, hence utilize more time. 
These computations were performed (sequentially) on an x86\_64 machine
under GNU/ Linux. The method is also effective for $I_{10}$ (4D), 
but needs parallelization for $I_{11}$ (5D).

{\color {black}
As part of our discussion, let's compare the performance of single 
and double extrapolations for the same problem.
To conclude initially, we observed that double extrapolation may
be justified to achieve better accuracy.
} 
{\color {black} This is illustrated for $I_{00}$ with $s = 7.$ 
The integrations are done by iterated adaptive integration using
the routine {\sc Dqagse} from {\sc Quadpack}~\cite{pi83,jocs11} with a maximum of
50 subdivisions in each coordinate direction,
relative error tolerance $10^{-9}$ in the outer coordinate and $5\times 10^{-10}$ in subsequent directions.

A double extrapolation with $\varepsilon = 1/1.15^\ell, ~\varrho = 1/1.2^k, 25 \le \ell \le 30, ~10 \le k \le 20$ yields for $C_{-2}, C_{-1}, C_0$ and $C_1$ at the $5^{th}$ linear system: 
0.1250000066$\pm$7.5e-08, 0.2687828385$\pm$1.7e-05, \\ 
-2.7440527816$\pm$1.5e-03, -4.5561503884.
The time for all six $\varepsilon$-cycles is 18.1\,seconds. 
{\color {black} For the same problem, a single extrapolation with ~$\varepsilon = 1/1.15^\ell, \ell \ge 25, ~~\varrho = 0$ ~~gives as the best result observed (at system 4):
0.1249996422$\pm$4.7e-06, 0.2688439263$\pm$7.8e-04, -2.7476802260$\pm$4.8e-02, -4.4743466989. 
In order to compare the time with the double extrapolation,
the result for $C_{-2}, C_{-1}, C_0, C_1$ at system 5 is:
0.1250017027$\pm$2.1e-06, 0.2683866114$\pm$4.6e-04, -2.7074720045$\pm$1.5e-02, -6.2248242790, with
a total time (for integrations 1 to 6) of 4.9 seconds.
}
}
\begin{figure}[b]
\includegraphics[width=0.5\linewidth]{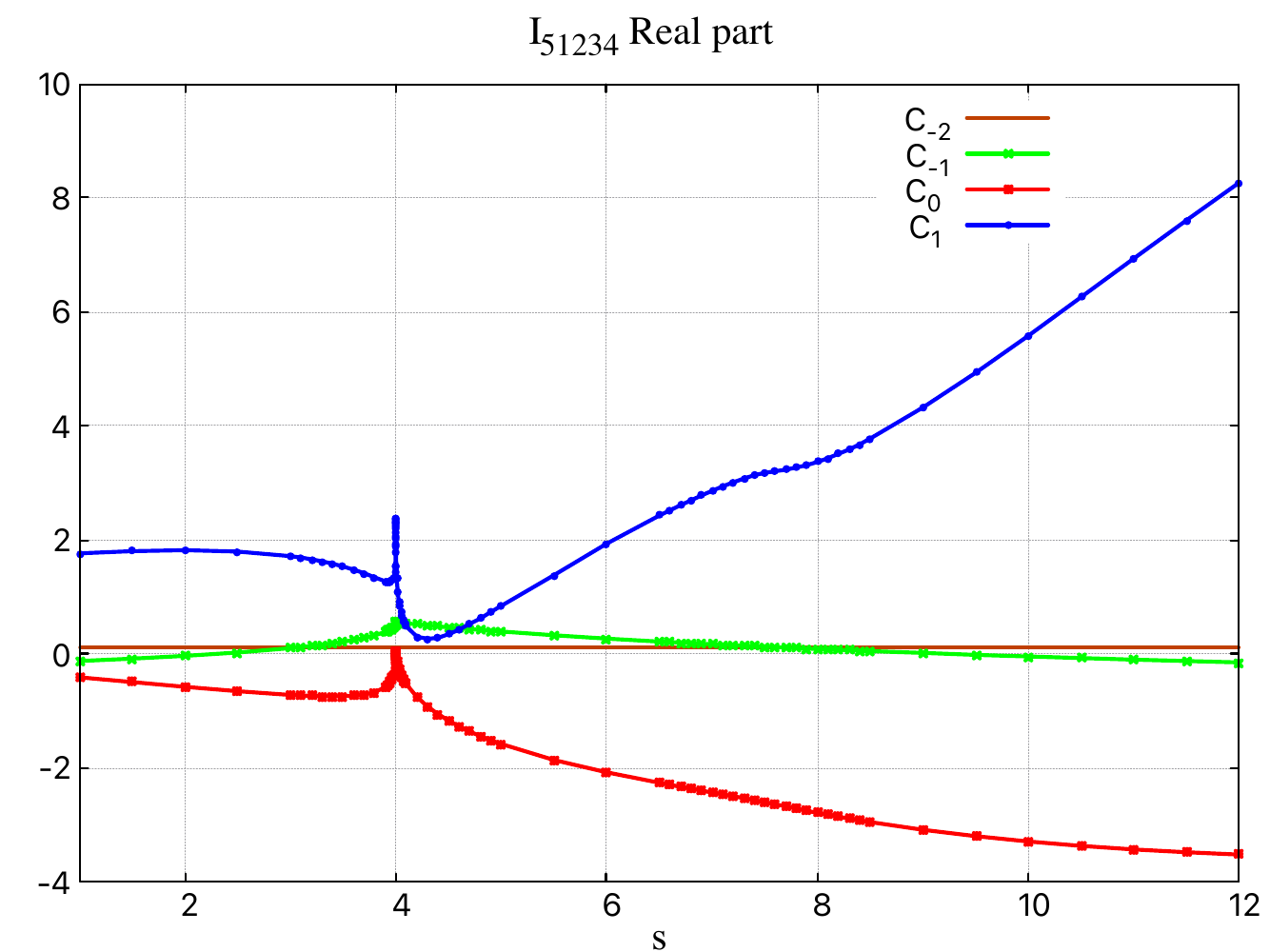}
\includegraphics[width=0.5\linewidth]{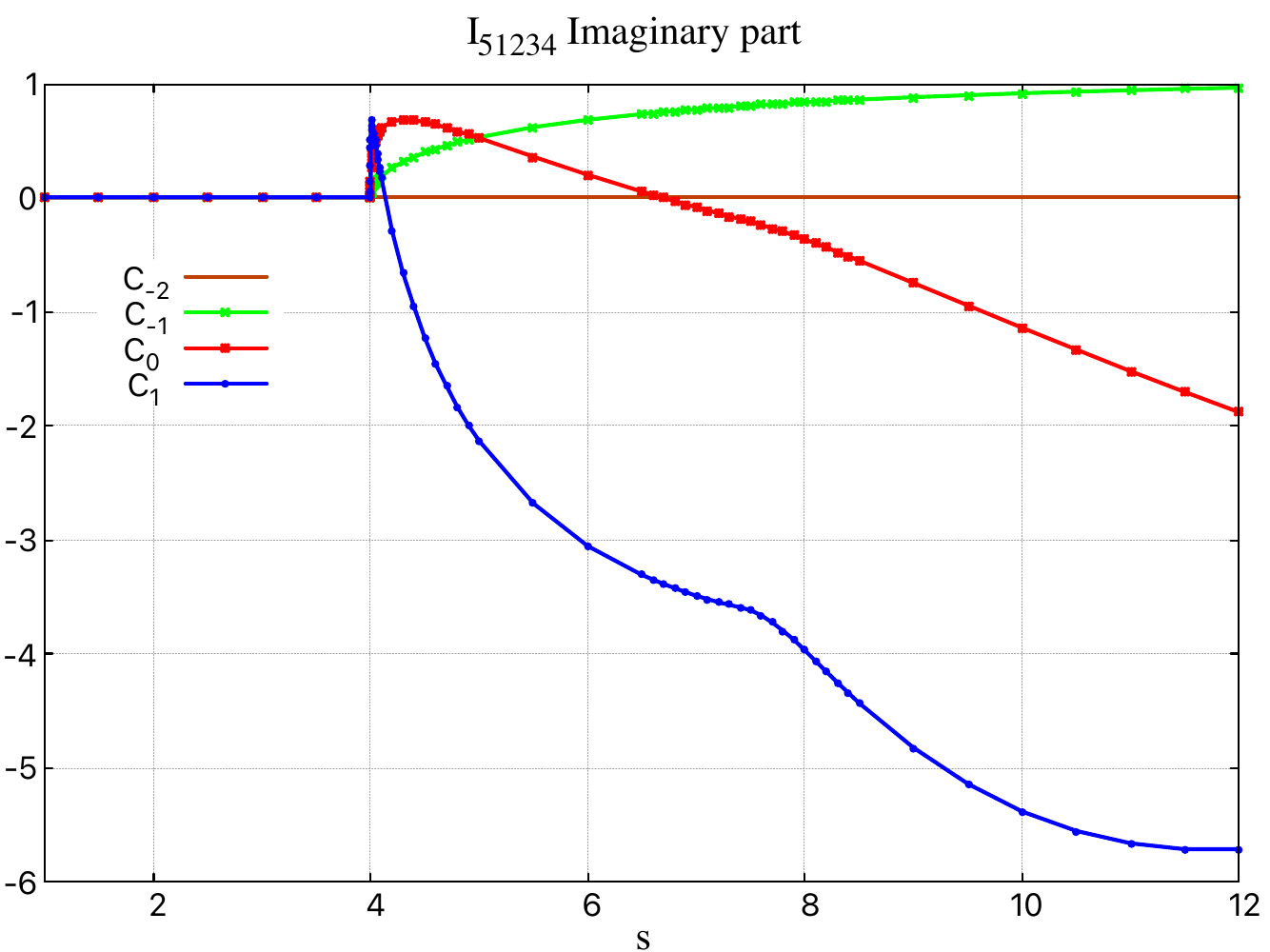}
\caption{Real and imaginary part of integral ${\mathcal I}_{51234}$
as a function of $s$ {\color {black} with DE integrattoin}}
\end{figure}

{\color {black} Finally, graphs} are displayed in Figure~2 for the real and imaginary part of the total integral 
${\mathcal I}_{51234}$ of Eq.\,(2), where the data were obtained using DE
integration and single extrapolation {\color {black} with regard to $\rho$ after an {\color{black}expansion in} $\varepsilon$}{\color {black} ~\cite{ptep24a}}.
~~\\

\section{Conclusions}
Whereas symbolic or symbolic/numerical calculations are performed for some challenging
problems using existing software packages, we focus on the development of 
fully numerical methods for the evaluation of Feynman loop integrals. 
The integration
strategies adhere to black-box approaches for generating
an approximation, assuming little or no knowledge of the problem, apart from the
specification of the integrand function. 
We demonstrated efficient strategies based on lattice rules evaluated on GPUs and double-exponential integration, as well as
iterated adaptive integration.

\ack
We acknowledge the support by JSPS KAKENHI Grant Number JP20K11858,
JP20K03941 and JP21K03541, and the National Science Foundation
Award Number 1126438 that funded work on multivariate integration.

\section*{References}
\bibliographystyle{iopart-num}
\bibliography{./bib3}

\providecommand{\newblock}{}
\begin{thebibliography}{10}
\expandafter\ifx\csname url\endcsname\relax
  \def\url#1{{\tt #1}}\fi
\expandafter\ifx\csname urlprefix\endcsname\relax\def\urlprefix{URL }\fi
\providecommand{\eprint}[2][]{\url{#2}}

\bibitem{talk-ACAT2022}
de~Doncker E, Yuasa F, Ishikawa T and Kato K 2022 Loop integral computation in
  the {E}uclidean or physical kinematical region using numerical integration
  and extrapolation {\em Proceedings of ACAT 2022\/} {T}o appear,
  https://indico.cern.ch/event/1106990/contributions/4997231/

\bibitem{nakanishi57}
Nakanishi N 1957 {\em Prog.\ Theor.\ Phys.\/} {\bf 17} 401--418

\bibitem{kinoshita74A}
Cvitanovi\'c P and Kinoshita T 1974 {\em Phys.\ Rev.\/} {\bf D10} 3978

\bibitem{kinoshita74B}
Cvitanovi\'c P and Kinoshita T 1974 {\em Phys.\ Rev.\/} {\bf D10} 3991

\bibitem{kinoshita74C}
Cvitanovi\'c P and Kinoshita T 1974 {\em Phys.\ Rev.\/} {\bf D10} 4007

\bibitem{sloan}
Sloan I and Joe S 1994 {\em Lattice Methods for Multiple Integration\/} (Oxford
  University Press)

\bibitem{sidi03}
Sidi A 2003 {\em Practical Extrapolation Methods - Theory and Applications\/}
  (Cambridge Univ. Press) iSBN 0-521-66159-5

\bibitem{sidi06}
Sidi A 2005 {\em Math. Comp.\/} {\bf 75} 327--343

\bibitem{mori78}
Mori M 1978 {\em Publ. RIMS Kyoto Univ.\/} {\bf 14} 713--729
  https://doi.org/10.2977/prims/1195188835

\bibitem{takahasi74}
Takahasi H and Mori M 1974 {\em Publications of the Research Institute for
  Mathematical Sciences\/} {\bf 9} 721--741

\bibitem{sugihara97}
Sugihara M 1997 {\em Numerische Mathematik\/} {\bf 75} 379--395

\bibitem{iccsa22}
de~Doncker E and Yuasa F 2022 {\em Springer Lecture Notes in Computer Science
  (LNCS)\/} {\bf 13378} 388--405

\bibitem{pi83}
Piessens R, de~Doncker E, {\"U}berhuber C~W and Kahaner D~K 1983 {\em QUADPACK,
  A Subroutine Package for Automatic Integration\/} ({\em Springer Series in
  Computational Mathematics\/} vol~1) (Springer-Verlag)

\bibitem{parintweb}
ParInt {http://www.cs.wmich.edu/parint}

\bibitem{csci19}
Olagbemi O~E and de~Doncker E 2019 Scalable algorithms for multivariate
  integration with {P}ar{A}dapt and {CUDA} {\em Proc. 2019 Int. Conf. on Comp.
  Science and Comp. Intelligence\/} (IEEE Computer Society)
  https:doi.org/10.1109/CSCI49370.2019.00093

\bibitem{jocs11}
de~Doncker E, Fujimoto J, Hamaguchi N, Ishikawa T, Kurihara Y, Shimizu Y and
  Yuasa F 2011 {\em Journal of Computational Science (JoCS)\/} {\bf 3} 102--112
  {DOI}:10.1016/j.jocs.2011.06.003

\bibitem{ptep24a}
de~Doncker E, Ishikawa T, Kato K and Yuasa F 2024 {\em Prog. Theor. Exp.
  Phys.\/} {h}ttps://doi.org/10.1093/ptep/ptae122

\bibitem{almulihi17}
Almulihi A and de~Doncker E 2017 Accelerating high-dimensional integration
  using lattice rules on {GPU}s {\em Proc. 2017 Int. Conf. on Computational
  Science and Computational Intelligence (CSCI'17)\/} (CPS IEEE)

\bibitem{nuyens06}
Nuyens D and Cools R 2006 {\em Math. Comp.\/} {\bf 75} 903--920

\bibitem{nuyens06a}
Nuyens D and Cools R 2006 {\em Journal of Complexity\/} {\bf 22} 4--28

\bibitem{brezinski80}
Brezinski C 1980 {\em Numerische Mathematik\/} {\bf 35} 175--187

\bibitem{wynn56}
Wynn P 1956 {\em Mathematical Tables and Aids to Computing\/} {\bf 10} 91--96

\bibitem{cpp10}
de~Doncker E, Fujimoto J, Hamaguchi N, Ishikawa T, Kurihara Y, Ljucovic M,
  Shimizu Y and Yuasa F 2010 Extrapolation algorithms for infrared divergent
  integrals arXiv:1110.3587

\bibitem{acat11}
de~Doncker E, Yuasa F and Kurihara Y 2012 {\em Journal of Physics: Conf.
  Ser.\/} {\bf 368} doi:10.1088/1742-6596/368/1/012060

\bibitem{yuasa-JPS22}
Yuasa F, de~Doncker E, Ishikawa T, Kato K, Daisaka H and Nakasato N 2022
  Numerical method for {F}eynman integrals: {E}lectroweak high-order correction
  calculation by {DCM IV} {F}all Meeting of the Physical Society of Japan,
  8aA133-6

\bibitem{bulirsch64}
Bulirsch R 1964 {\em Numerische Mathematik\/} {\bf 6} 6--16

\end{thebibliography}
\end{document}